\documentclass[10pt,conference]{apmc2022}
\ifCLASSINFOpdf
  \usepackage[pdftex]{graphicx}
\else
  \usepackage[dvips]{graphicx}
\fi

\usepackage{amsmath,amssymb,amsfonts}
\usepackage{cite}

\usepackage{threeparttable}
 \usepackage{array} 
\usepackage{soul}
\begin{document}
%\title{Study of Reducing Signal Fluctuation on\\ Reconfigurable Porous Surface Wave Platform}
\title{On Surface Wave Propagation Characteristics of\\Porosity-Based Reconfigurable Surfaces}

%\author{\IEEEauthorblockN{Michael Shell}
%\IEEEauthorblockA{School of Electrical and\\Computer Engineering\\
%Georgia Institute of Technology\\
%Atlanta, Georgia 30332--0250\\
%Email: http://www.michaelshell.org/contact.html}
%\and
%\IEEEauthorblockN{Homer Simpson}
%\IEEEauthorblockA{Twentieth Century Fox\\
%\and
%\IEEEauthorblockN{James Kirk\\ and Montgomery Scott}
%\IEEEauthorblockA{Starfleet Academy\\
%San Francisco, California 96678--2391\\



\author{\IEEEauthorblockN{Zhiyuan Chu, Kai-Kit Wong, and Kin-Fai Tong}
\vspace{0.75ex}
\IEEEauthorblockA{
{\em Department of Electronic and Electrical Engineering, University College London}\\
{\em Torrington Place, WC1E 7JE, London, United Kingdom}\\
E-mail: $\{\rm zhiyuan.chu.18,kai\text{-}kit.wong,k.tong\}@ucl.ac.uk$}
%kai-kit.wong@ucl.ac.uk\\
%k.tong@ucl.ac.uk}
}

\maketitle 

\begin{abstract}
Reconfigurable surfaces facilitating energy-efficient, intelligent surface wave propagation have recently emerged as a technology that finds applications in many-core systems and 6G wireless communications. In this paper, we consider the porosity-based reconfigurable surface where there are cavities that can be filled on-demand with fluid metal such as Galinstan, in order to create adaptable channels for efficient wave propagation. We aim to investigate the propagation phenomenon of signal fluctuation resulting from the diffraction of discrete porosity and study how different porosity patterns affect this phenomenon. Our results cover the frequency range between $21.7{\rm GHz}$ and $31.6{\rm GHz}$ when a WR-34 waveguide is used as the transducer.
\end{abstract}


\IEEEpeerreviewmaketitle

\section{Introduction}
Surface waves have recently gained some attention for their efficient radio propagation and surface wave communications has been proposed in many different applications such as on-body communications \cite{berkelmann2021antenna}, network-on-chip (NoC) for many-core systems \cite{karkar2022thermal}, and etc. There has also been an upsurge of research efforts of utilizing large intelligent surfaces, mostly referred to as reconfigurable intelligent surface (RIS), for 6G wireless communications \cite{tang2019programmable,di2020reconfigurable}. While the current activities concentrate on performing intelligent reflections directing to the users of interest, \cite{wong2020vision} further discusses the potential benefits of adopting surface waves to control interference and improve propagation efficiency in wireless communication networks.

An advantage of surface wave communications is that reconfigurable surfaces such as the one proposed in \cite{chu2021enhancing} are possible so that dynamic low-loss channels or pathways can be created on-demand to route the signals in any desirable way. In \cite{chu2021enhancing}, such reconfigurable surface is achieved by a porous surface in which the cavities can be filled with conductive liquid to form isolated pathways via digitally controlled pumps. There is however lack of understanding of how the porosity pattern affects the propagation performance of the surface.

Motivated by this, our aim is to investigate the propagation characteristics of the porosity-based reconfigurable surface by considering different porosity patterns and sheds light on what makes a better surface in terms of signal fluctuation, which is an undesirable phenomenon resulting from the diffraction of discrete porosity. Our results demonstrate that a surface with interleaved cavities performs the best in reducing fluctuation.

\begin{figure}[]
\centering
\includegraphics[width=9cm]{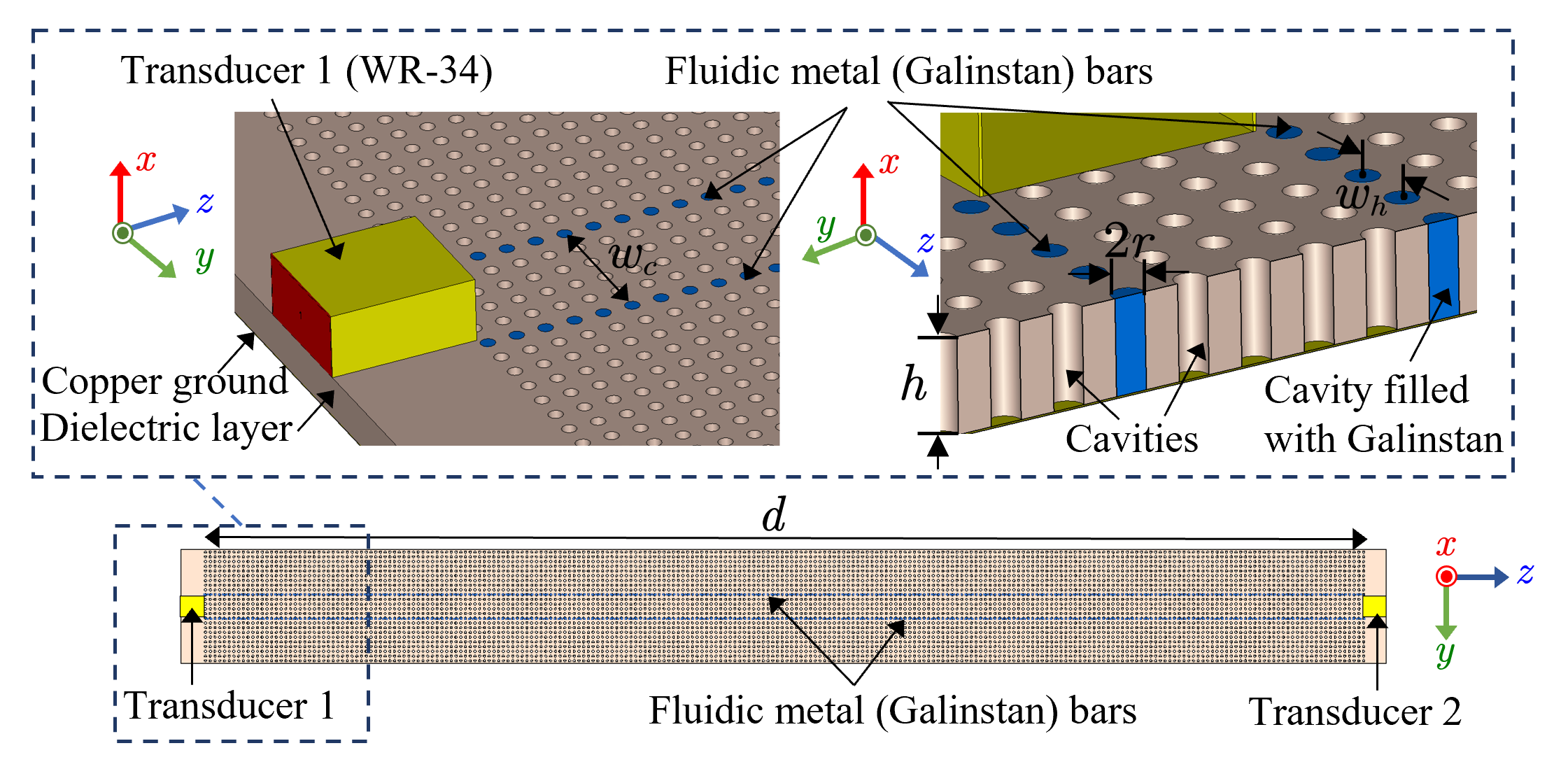}
\caption{The reconfigurable surface with cavities fillable by fluid metal.}\label{fig:model}
\vspace{-3mm}
\end{figure}

\begin{figure}[]
\centering
\includegraphics[width=9cm]{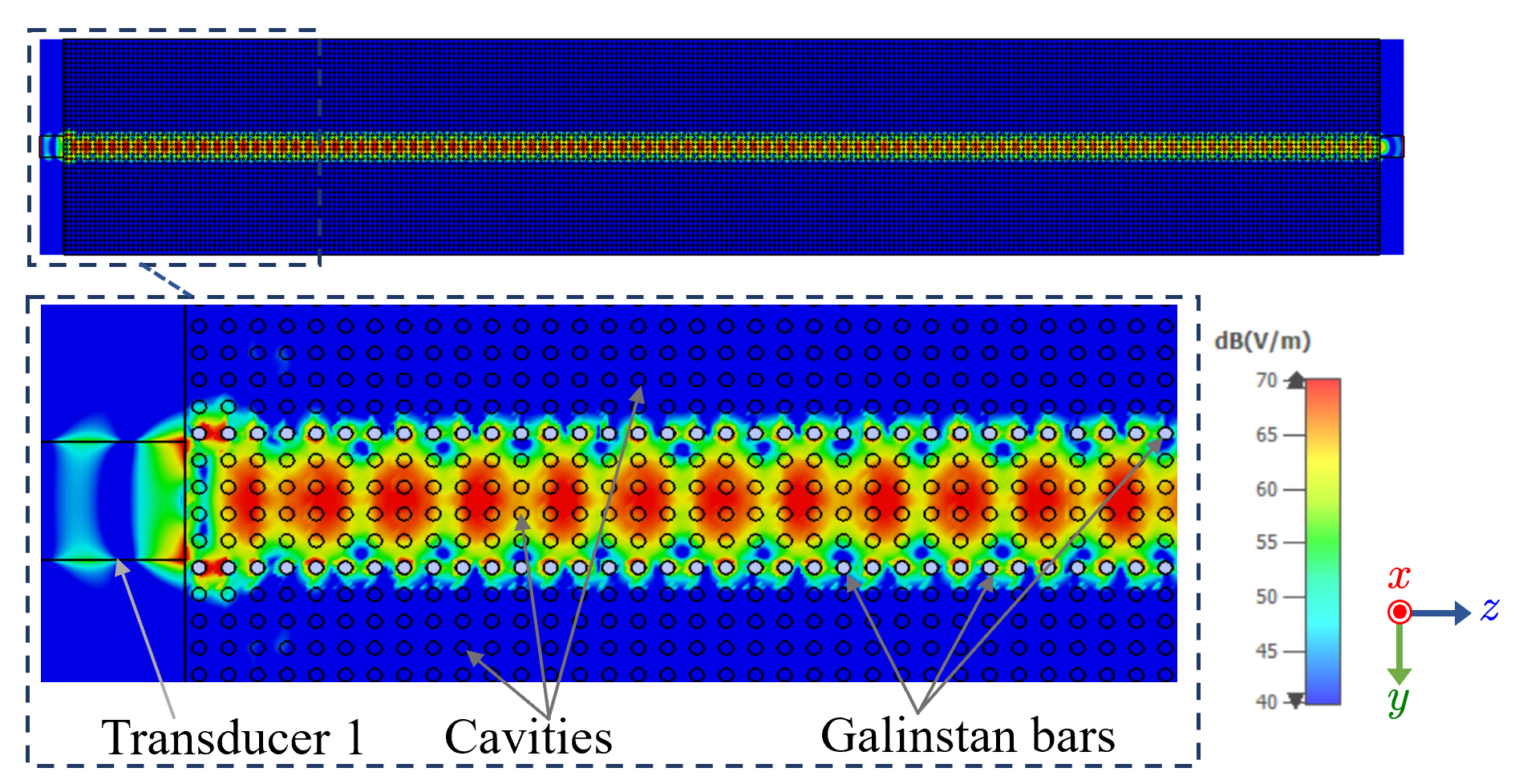}
\caption{Illustration of an isolated channel on the reconfigurable surface.}\label{fig:simulation}
\vspace{-3mm}
\end{figure}

\renewcommand{\arraystretch}{1.15} 

\begin{table}[t]
\begin{center}
\begin{tabular}{r|c}

{Parameter} & {Value}\\
\hline\hline
\mbox{channel width}, $w_c$ & $9{\rm mm}$\\
\hline
\mbox{radius of the bar and cavity}, $r$ & $0.5{\rm mm}$\\
\hline
\mbox{thickness of dielectric layer}, $h$ & $2.85{\rm mm}$\\
\hline
\mbox{channel propagation distance}, $d$ & $2000{\rm mm}$\\
\hline
\mbox{inductive surface impedance}, $X_s$ & $j270\Omega$\\
\hline
\mbox{relative permittivity of dielectric layer}, $\varepsilon_r$ & $2.1$\\
\hline
\mbox{conductivity for Galinstan}, $\sigma_{\rm g}$ & $3.46\times 10^6~{\rm Sm}^{-1}$\\
\hline
\mbox{conductivity for copper}, $\sigma_{\rm c}$ & $59.6\times 10^6~{\rm Sm}^{-1}$\\
\hline
\mbox{operating frequency}, $f$ & $26{\rm GHz}$\\
\hline
\mbox{transducer (WR-34) frequency band}, $f_b$ & $22-33{\rm GHz}$\\
\end{tabular}
\end{center}
\caption{Key parameters of the surface.}\label{tab:1}
\vspace{-5mm}
\end{table}
\begin{figure}[]
\begin{threeparttable}
\centering
\includegraphics[width=8.5cm]{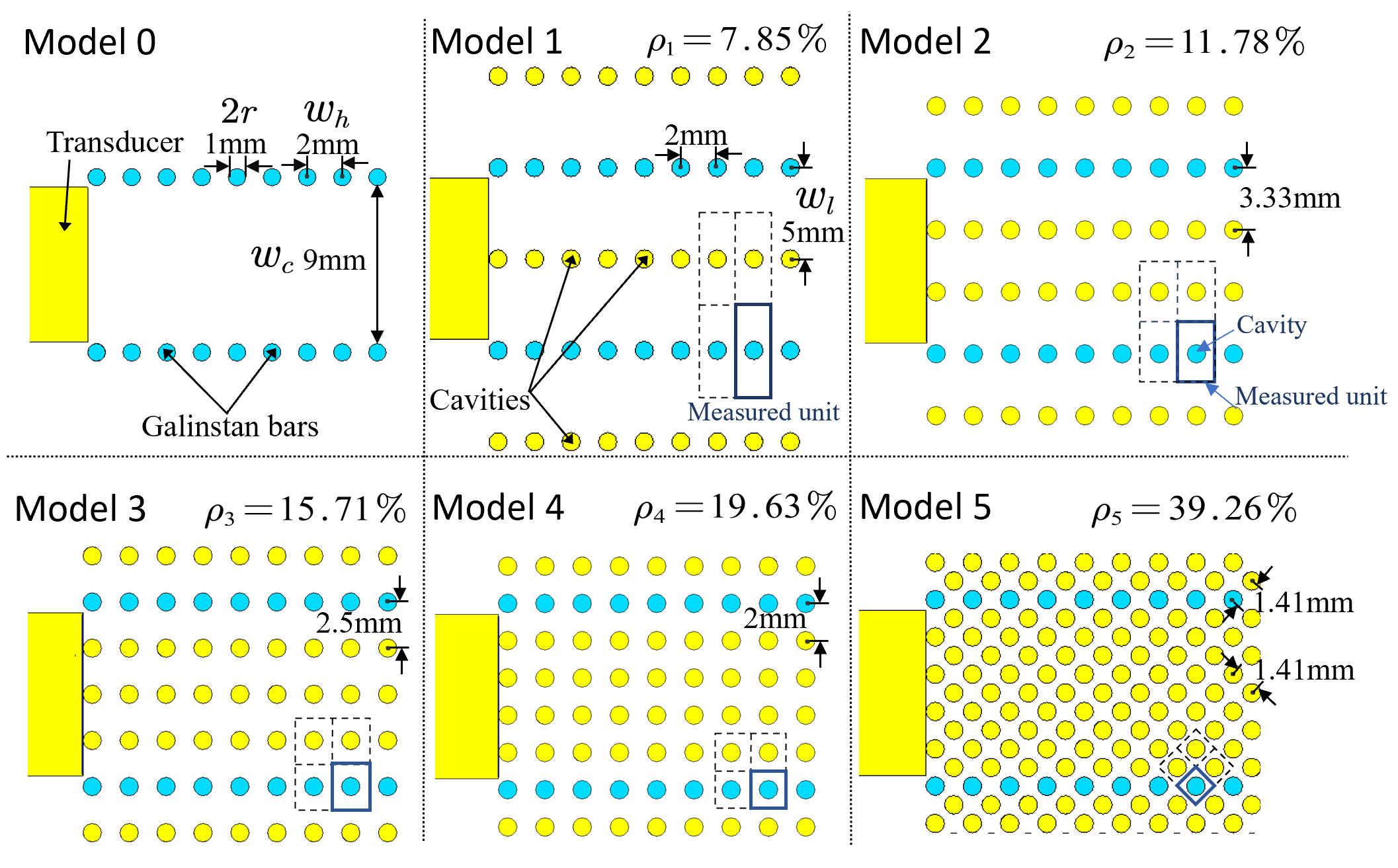}
 \begin{tablenotes}
        \footnotesize
        \item[*] Porosity is defined as $\rho =S_{\mathrm{cavity}}/S_{\mathrm{measured}\text{-}\mathrm{unit}}$.
      \end{tablenotes}
    \end{threeparttable}
\caption{Models with different cavity distribution densities, or porosity patterns.}\label{fig:different_model}
\end{figure}

 \begin{figure*}[]
\centering
\includegraphics[width=18.5cm]{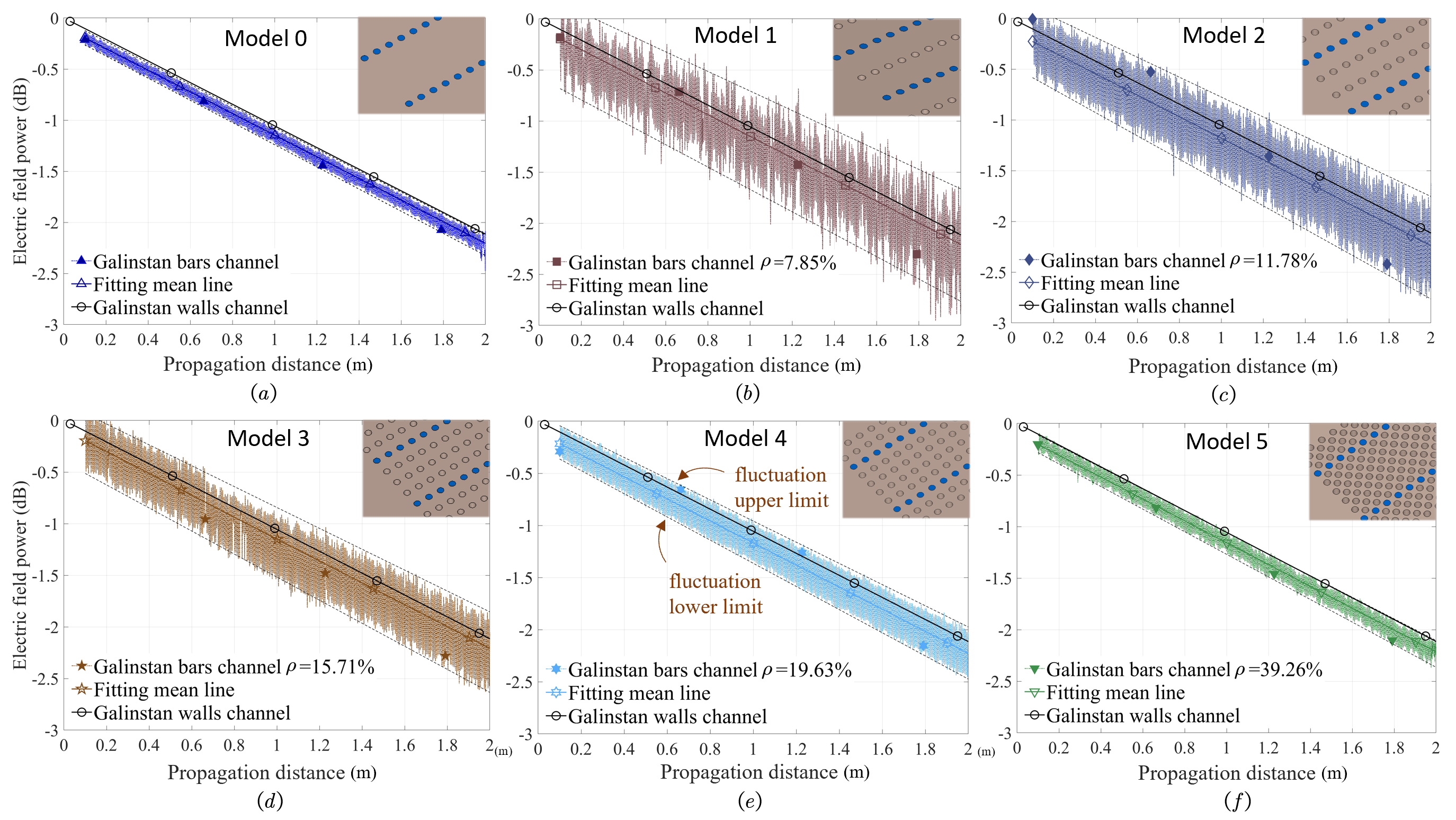}
\caption{The electric field power (${\rm dB}$) results for the reconfigurable surface with $ (a) $ only the Galinstan bars channel, $ (b) $ the Galinstan bars channel with porosity $\rho=7.85\%$,  $ (c) $ $\rho=11.78\%$, $ (d) $ $\rho=15.71\%$, $ (e) $ $\rho=19.63\%$, and $ (f) $ interleaved cavities with $\rho=39.26\%$.}\label{fig:different_results}
\end{figure*}

\section{Porosity-Based Reconfigurable Surface}
As shown in Fig.~\ref{fig:model}, the reconfigurable surface under consideration has evenly distributed cavities where conductive fluid such as Galinstan can be injected on demand. When Galinstan fills up two rows of cavities to form a channel, isolated surface wave propagation can be achieved, see the results in Fig.~\ref{fig:simulation}. The parameters of the model are given in TABLE \ref{tab:1}. 

Five models of porosity pattern, referred to as Model 1 to Model 5, as illustrated in Fig.~\ref{fig:different_model}, are considered. The parameter, $w_l$, in the model can be changed to specify different values of porosity which is defined as $\rho=S_{\rm cavity}/S_{\rm measured\text{-}unit}$ where $S_{\rm cavity}$ denotes the top area of each cavity and $S_{\rm measured\text{-}unit}$ is the top area of the measured unit (i.e., the rectangular area marked in the figure). With porosity $\rho$, the effective relative permittivity of the surface changes and can be found as \cite{liu2016general} 
\begin{equation}
\varepsilon_r^{\rm eff}=\frac{\varepsilon_r\left[1+3\varepsilon_r+3\rho(1-\varepsilon_r)\right]}{1+3\varepsilon_r +\rho(\varepsilon_r-1)},
\end{equation}
where $\varepsilon_r$ denotes the original relative permittivity. The surface impedance then can be obtained by \cite{barlow1953surface} 
\begin{equation}
X_s=2\pi f\mu_0\left(\frac{\varepsilon_r^{\rm eff}-1}{\varepsilon_r^{\rm eff}}h+\frac{\Delta}{2}\right).
\end{equation}
where $\Delta$ denotes the skin depth of the metal sheet, e.g., the copper ground plane, $h$ is the dielectric layer thickness, $f$ is the operating frequency and $\mu_0=4\pi\times 10^{-7}~{\rm Hm}^{-1}$ is the permeability of free space.

To have the maximum excitation efficiency, $X_s$ should be set to $j270{\rm \Omega}$ \cite{Wan-TAP2019}. This is achieved by adjusting the dielectric layer thickness $h$ appropriately to match the different porosities for a fixed surface impedance. The corresponding values of the parameters are presented in TABLE \ref{tab:2}. 

\begin{table}[t]
\begin{center}
\begin{tabular}{@{}cccccc@{}}
 \hline

\begin{tabular}[c]{p{0.6cm}<{\centering}@{}c@{}}Model\\ \\ \\ \end{tabular} & \begin{tabular}[c]{p{1.2cm}<{\centering}@{}c@{}}Longitudinal\\separation\\$w_l$(${\rm mm}$)\end{tabular} & 
\begin{tabular}[c]{p{1cm}<{\centering}@{}c@{}}Horizontal \\ separation\\$w_h$(${\rm mm}$)\end{tabular} & \begin{tabular}[c]{p{0.7cm}<{\centering}@{}c@{} }Porosity\\ \\ $\rho $(\%)\end{tabular} & \begin{tabular}[c]{p{1.1cm}<{\centering}@{}c@{}}Effective \\ permittivity\\ $\varepsilon _r^{\mathrm{eff}}$\end{tabular} & \begin{tabular}[c]{@{}c@{}}Thickness\\ \\ $h$(${\rm mm}$)\end{tabular} \\  \hline
0&--&--&0& 2.10& 2.50\\ 
1&5.00&2&7.85&2.00&2.63\\ 
2&3.33&2&11.78&1.95&2.69 \\  
3&2.50&2&15.71&1.91&2.77\\ 
4&2.00&2&19.63&1.86&2.85\\  
%6&1.67&2&23.56&1.81&2.93\\  
5&1.41&2&39.26&1.63&3.40\\  \hline
\end{tabular}
\end{center}
\caption{The parameter variations in the comparative models.}\label{tab:2}
\vspace{-5mm}
\end{table}

\section{Simulation Results}
To evaluate the models, electromagnetic (EM) simulations were performed by CST Studio Suite 2020. PTFE ($\varepsilon_r=2.1$, $\tan\delta = 0.0002@10{\rm GHz}$) was used as the dielectric layer and a rectangular waveguide WR-34 was used as the transducer with a height of $2.8{\rm mm}$ and width of $9.6{\rm mm}$ for surface wave transmission in this design. Fig.~\ref{fig:different_results} demonstrates a set of electric field distribution results in ${\rm dB}$ over the surface inside the straight channel localized by the Galinstan bars. The results show much fluctuation caused by standing wave reflection and diffraction from the cavities. To facilitate comparison with the case without the cavities shown in Fig.~\ref{fig:different_results}(a), we plot the numerical means of the models. The radius $r$ of the cavities is referenced to the previous work in \cite{chu2021enhancing} that is the paper we published before about determination of the radius.

In TABLE \ref{tab:3}, we provide the standard deviation (SD) of the electric field fluctuation, $\sigma$, and the path loss of each model. As can be observed, the fluctuation SD, $\sigma$, decreases gradually from $0.297$ to $0.070$ from Model 1 to Model 5 as the surface porosity increases from $7.85\%$ to $39.26\%$, suggesting that a surface with denser cavities helps reduce the signal fluctuation, approaching closer to Model 0 which has a signal fluctuation SD of just $0.052$. Additionally, the discrepancy in path losses in different models is below $0.05{\rm dB}$ in a $2000{\rm mm}$ ($173.3\lambda$ at $26{\rm GHz}$) propagation distance, indicating that potential loss caused by the porosity is negligible if the surfaces are kept at the same surface impedance.

\begin{table}[]
\begin{center}
\begin{threeparttable}
\begin{tabular}{@{}cccccc@{}}
 \hline
\begin{tabular}[c]{p{1.2cm}<{\centering}@{}c@{}}Model\\ \\ \end{tabular} & 
\begin{tabular}[c]{p{1.2cm}<{\centering}@{}c@{}}Porosity \\ $\rho $(\%)\end{tabular} &
\begin{tabular}[c]{p{2.3cm}<{\centering}@{}c@{}}Fluctuation SD\\$\sigma$\end{tabular} &  \begin{tabular}[c]{p{1.5cm}<{\centering}@{}c@{}}Path loss\\ $L$(${\rm dB}$ /${\rm m}$)\end{tabular} & \\  \hline

0&0& 0.052& 1.10\\ 
1&7.85&0.297&1.12\\ 
2&11.78&0.265&1.13 \\  
3&15.71&0.221&1.10\\ 
4&19.63&0.131&1.11\\  
5&39.26&0.070&1.12\\  \hline
\end{tabular}

 \begin{tablenotes}
        \footnotesize
        \item[1] SD is the standard deviation, $\sigma =\sqrt{\frac{\sum{\left( x_i-\mu \right) ^2}}{n}}.$  
        \item[2] $n$ is the number of data samples, $x_i$ is the value of each sample and $\mu$ is the value of the local mean.
        \item[3] Path loss $L$ is measured using the numerical mean line in each model.
      \end{tablenotes}
    \end{threeparttable}

\end{center}
\caption{Signal fluctuation and path loss of the models.}\label{tab:3}
\vspace{-8mm}
\end{table}

%Particularly, It is noticed in Model $5$ that the use of proposed reconfigurable interleaving porous surface with an even $1.41{\rm mm}$ cavity separation could be effectively instrumental to the subduction of $E$-field fluctuation. More cavity density means more combination possibilities and these staggered cavities filled with Galinstan could provide a more flexible pathway generation containing turns with different angles and positions for the realization of controllable surface waves in our follow-up study. 

\begin{figure}[]
\centering
\includegraphics[width=9cm]{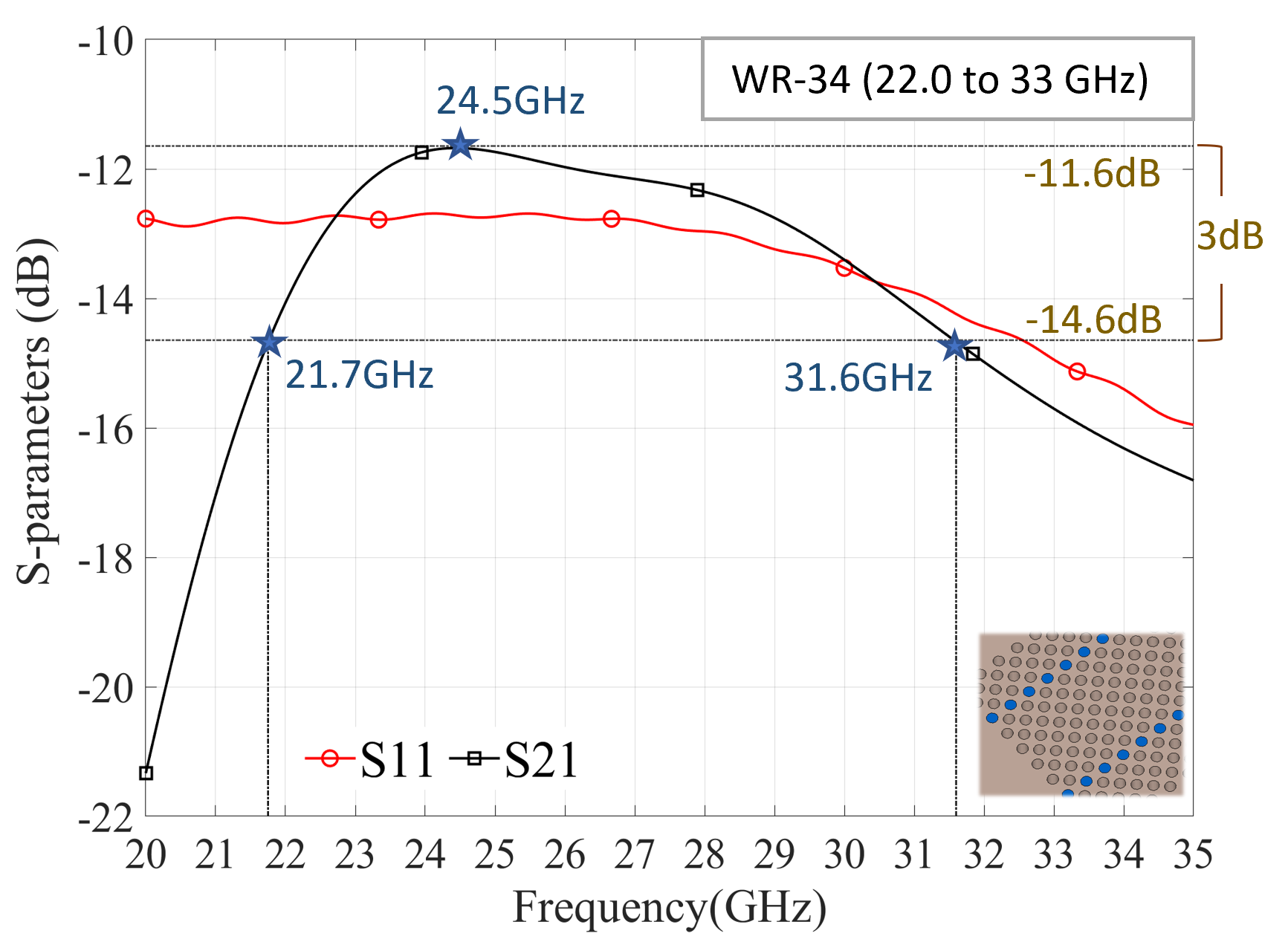}
\caption{The S$11$ and S$21$ simulation results for the interleaving porous surface operating in a wide frequency band from $21.7{\rm GHz}$ to $31.6{\rm GHz}$.}\label{fig:S11_S21}

\end{figure}

We conclude this section by studying the wideband performance of the reconfigurable surface with interleaved cavities, i.e., Model 5. The S11 and S21 results over the frequency from $20{\rm GHz}$ to $35{\rm GHz}$ are presented in Fig.~\ref{fig:S11_S21}. The results reveal that the peak, i.e., the optimum frequency occurs at $24.5{\rm GHz}$ with S21 of $-11.6{\rm dB}$. Moreover, the half-power $3$-${\rm dB}$ bandwidth is measured to be located from $21.7{\rm GHz}$ to $31.6{\rm GHz}$ which may be only limited by the cut-off frequency of the transducer at $22{\rm GHz}$ and $33{\rm GHz}$. In summary, the porosity-based reconfigurable surface works over a wide band although it still needs to keep an appropriate surface impedance by adjusting the thickness to match the different working frequencies of the transducers on-demand.

\section{Conclusion}
This paper investigated the impact of porosity on reconfigurable surfaces through EM simulations. The signal fluctuation phenomenon was discussed while different porosity patterns were investigated to understand how porosity would affect the performance. Our results illustrated that the signal fluctuation could be much reduced with a denser porosity pattern and the porosity-based reconfigurable surface demonstrated promising performance over a wide bandwidth.

%\vspace{10mm}
%\underline{The 2022 Asia-Pacific Microwave Conference}
%\underline{Paper Submission Deadline : June 17, 2022 }
%\underline{Held in Yokohama, Japan : Nov. 29 - Dec. 2, 2022}
\end{document}